\documentclass{WileyMSP_template}
\usepackage{amssymb}
\usepackage{xcolor}
\usepackage{parskip}

\begin{document}

\pagestyle{fancy}
%\rhead{\includegraphics[width=2.5cm]{vch-logo.png}}

\title{Shaping a Surface Microdroplet by Marangoni Forces along a Moving Contact Line of Four Immiscible Phases}

\maketitle

% Author: Please give full first and last names for authors and include * after the name of all corresponding authors

\author{Haichang Yang}
\author{Binglin Zeng}
\author{Qiuyun Lu}
\author{Yaowen Xing}
\author{Xiahui Gui}
\author{Yijun Cao}
\author{Ben Bin Xu*}
\author{Xuehua Zhang*}

% Dedication
%\dedication{}

% Affiliations: Please provide adacemic titles (Prof. or Dr.) for all authors where applicable, and include an institutional email address for all corresponding authors
\begin{affiliations}

Dr. Haichang Yang, Dr. Binglin Zeng, Dr. Qiuyun Lu, Prof. Xuehua Zhang*\\
Department of Chemical and Materials Engineering, University of Alberta, Edmonton, T6G 1H9, Alberta, Canada\\
Email Address: xuehua.zhang@ualberta.ca

Dr. Haichang Yang, Prof. Yaowen Xing, Prof. Xiahui Gui, Prof. Yijun Cao\\
National Engineering Research Center of Coal Preparation and Purification, China University of Mining and Technology, Xuzhou, 221116, Jiangsu, China\\

Prof. Yijun Cao\\
School of Chemical Engineering and Technology, Zhengzhou University, Zhengzhou, 450066, Henan, China\\

Prof. Ben Bin Xu*\\
Smart Materials and Surfaces Lab, Mechanical and Construction Engineering, Faculty of Engineering and Environment, Northumbria University, Newcastle upon Tyne, NE1 8ST, UK\\
Email Address: ben.xu@northumbria.ac.uk

\end{affiliations}


\begin{abstract}
The ability to transfer microdroplets between fluid phases offers numerous advantages in various fields, enabling better control, manipulation, and utilization of small volumes of fluids in pharmaceutical formulations, microfluidics, and lab-on-a-chip devices, single-cell analysis or droplet-based techniques for nanomaterial synthesis. This study focuses on the stability and morphology of a sessile oil microdroplet at the four-phase contact line of solid-water-oil-air during the droplet transfer from underwater to air. We observed a distinct transition in microdroplet dynamics, characterized by a shift from a scenario dominated by Marangoni forces to one dominated by capillary forces. In the regime dominated by Marangoni forces, the oil microdroplets spread in response to the contact between the water-air interface and the water-oil interface and the emergence of an oil concentration gradient along the water-air interface. The spreading distance along the four-phase contact line follows a power law relationship of $t^{3/4}$, reflecting the balance between Marangoni forces and viscous forces. On the other hand, in the capillarity-dominated regime, the oil microdroplets remain stable at the contact line and after being transferred into the air. We identify the crossover between these two regimes in the parameter space defined by three factors: the approaching velocity of the solid-water-air contact line ($v_{cl}$), the radius of the oil microdroplet ($r_o$), and the radius of the water drop ($r_w$). Furthermore, we demonstrate how to use the four-phase contact line for shaping oil microdroplets using a full liquid process by the contact line lithography. The findings in this study may be also applied to materials synthesis where nanoparticles, microspheres, or nanocapsules are produced by microdroplet-based techniques. 
\end{abstract}

\section{Introduction}

Controlling the behaviour of microdroplets is key to unlocking the technology potential in various processes. In nanomaterial synthesis, maintaining stable microdroplets with precise compositions and sizes is crucial to perform precise and controlled multistep reactions \cite{kaminski2017controlled,theberge2010microdroplets,yang2015compartmentalization,ma2022remodeling}, and to obtain desired product properties \cite{pearce2021,lee2015acceleration,kanike2023flow,billet2023hydrophilic}. For analytical chemistry and medical diagnostics, transferring microdroplets often enables the isolation and manipulation of small volumes of samples, leading to better sensitivity and accuracy in detecting specific compounds or analytes \cite{li2015splitting, qian2022situ, lu2017universal}. In other situations, it is desirable to mobilize microdroplets on surfaces or interfaces. For example, detaching water microdroplets from surfaces can also help prevent fouling and maintain surface self-cleaning \cite{lebedeva2011nano}. The ability to control the stability and morphology of microdroplets is often achieved by manipulating various factors, including interfacial tension, surface treatments or functional coatings\cite{han2022active,zhang2023lossless}, and external forces like electric fields or acoustic waves \cite{sinha2022patterning}. However, the potential of a moving interface between gas and an immiscible liquid has been rarely explored for controlling the morphology of microdroplets.

Extensive studies have been conducted on interfacial dynamics of a microdroplet in symmetric systems consisting of two approaching drops \cite{mettu2017charge, edwards2009effects}, and in asymmetric systems of a bubble approaching a drop \cite{li2021atomic, tabor2011homo}, or a drop approaching a solid surface \cite{liu2019probing}. When a bubble comes into proximity, the droplet immersed in a liquid medium may undergo deformation, forming a dimple \cite{allan1961approach}, pimple \cite{valkovska1999surfactants}, wimple \cite{clasohm2005wimple}, or ripple \cite{ajaev2008ripples}. The physicochemical properties of the liquid phase and the microdroplet play a significant role in either facilitating coalescence or stabilizing the droplets. The intervening liquid film between the droplet and the bubble is essential for droplet stability, which is determined by interfacial deformation under surface forces and the hydrodynamic drainage process. The microdroplet stability is related to the approaching velocity of the bubble surface, viscosity, electrolyte concentration, or pH value of the medium and the droplet size \cite{dagastine2006dynamic, hunter2001foundations, israelachvili2011intermolecular, russel1991colloidal}.

Various body forces are known to impact the dynamics and stability of droplets, such as inertial forces \cite{Aarts2005}, viscous forces \cite{brinkman1949calculation,yi2022local}, and capillary forces \cite{Aarts2005, upadhyay2020underwater}. Moreover, for a multiphase and multicomponent system out of equilibrium, the droplet behaviour is even more complicated by the presence of an interfacial tension gradient via Marangoni forces \cite{scriven1960marangoni,lu2017universal, zhang2021propelling, farzeena2023innovations}. In addition, for sessile droplets on a solid surface, the pinning effect along the contact line of droplets on surface structures may restrict the mobility of the droplets \cite{yan2019droplet}.

Distinct from both symmetric and asymmetric scenarios above, when an oil microdroplet on a solid substrate interacts with an approaching gas-water interface, a transient four-phase contact line can form. Specifically, the oil microdroplet is encapsulated in the host water drop as the water-air interface recedes, approaching the oil microdroplet. In our previous work, we observed the coalescence of an oil droplet and a water droplet on a hydrophobic substrate in a quasi-static system \cite{yu2019splitting}. The oil droplet split after the coalescence of these two drops, primarily governed by the interfacial tensions between the fluid phases and the viscosity of the oil droplet. However, there is a lack of quantitative understanding and control regarding the behaviour of surface microdroplets at a four-phase contact line of solid, gas, water, and oil. 

In this work, we will present the dynamic behaviour of a microdroplet at the four-phase contact line. Specifically, we have placed a small oil droplet inside a water drop and tuned the withdrawal of the water-air interface. Through these controlled experiments, we discovered a transition in the microdroplet dynamics between two distinct scenarios dominated by the Marangoni forces and capillary forces, respectively. This transition occurs as the four-phase contact line sweeps over the oil microdroplet on the substrate. The control parameters have been identified as the moving velocity of the water-air contact line, the radius of oil microdroplets, the radius of water droplets, and the oil type. We have demonstrated the utility of the four-phase contact line for shaping the oil microdroplets, showcasing the potential of using a full liquid process by the contact line lithography. The insights into the behavior of microdroplets in complex environments open new avenues for the design and control of droplet behaviors in diverse engineering processes. The understanding of the mechanisms that govern droplet behavior will be valuable for optimizing processes and tailoring them to specific needs, leading to advancements in various fields such as materials science, microfluidics, and environmental remediation.

\section{Results and Discussion}
\subsection{Two regimes in the microdroplet dynamics at the four-phase contact line}

\begin{figure}
  \centerline{\includegraphics[width=0.85\linewidth]{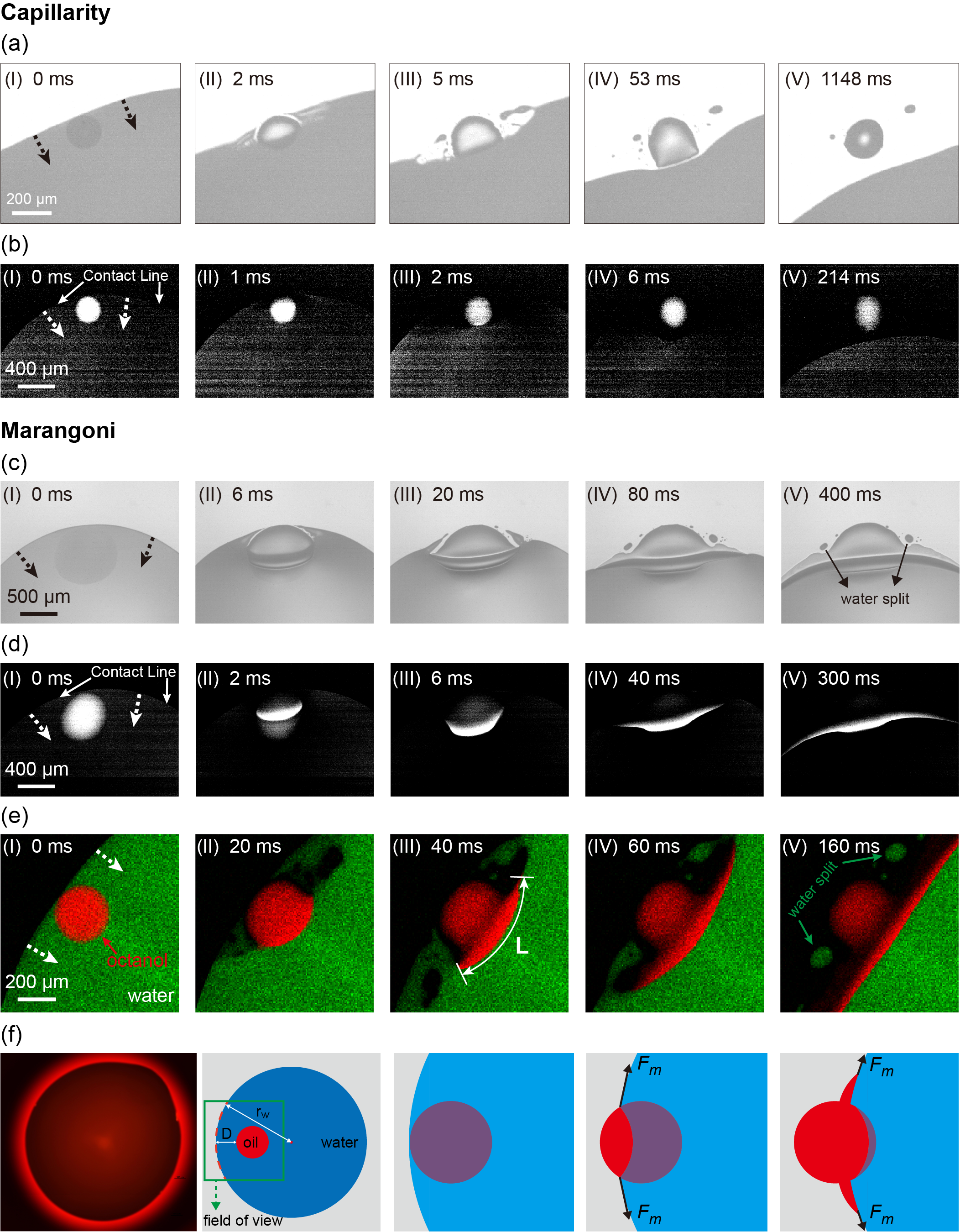}}
  \caption{(a) Snapshots captured by bright field optical microscope, where $r_0$ = 83 $\mu$m, $r_w$ = 2728 $\mu$m, and $v_{cl}$ = 0.36 mm/s; (b) Snapshots captured by fluorescence microscope, $r_o$ = 139 $\mu$m, $r_w$ = 1437 $\mu$m and $v_{cl}$ = 0.9 mm/s; (c) Snapshots captured by bright field optical microscope, where $r_0$ = 419 $\mu$m, $r_w$ = 1805 $\mu$m, and $v_{cl}$ = 0.07 mm/s; (d) Snapshots captured by fluorescence microscope, $r_o$ = 251 $\mu$m, $r_w$ = 1409 $\mu$m,  $v_{cl}$ = 0.01 mm/s; (e) Snapshots captured by confocal microscope, where the red and green color represent the octanol phase and water phase, respectively, $r_0$ = 145 $\mu$m, $r_w$ =  2109 $\mu$m,  and $v_{cl}$ =  0.01 mm/s; (f) The sketch of the dynamics of contact line on oil droplet. The videos record the process in (a-e) are attached in Supporting Information.} 
\label{CM}
\end{figure}

Figure \ref{CM} shows the typical snapshots of receding contact line over an octanol microdroplet with different v$_{cl}$ and $r_o$. Initially, the octanol microdroplet rests inside the water drop on the substrate. The term v$_{cl}$ is defined as the speed of the three-phase contact line of gas-water-solid before it reaches the octanol droplet. In the experiments, v$_{cl}$ was controlled by the flow rate of the injection pump. The $r_o$ is radius of contact line of octanol droplet. Here $t$ = 0 ms represents the time when solid-oil-water-gas four phase contact line forms, namely the moment when the three-phase contact line around the water drop just touches the oil droplet.

Figure \ref{CM}(a) shows the optical images of octanol droplet with radius of 83 $\mu$m and with v$_{cl}$ of 0.36 mm/s. The contact line quickly traverses the octanol microdroplet once the contact line meets the edge of microdroplet within 53 ms. Then, the contact line passes the droplet and remains fixed until the outer slowly moving parts catch up, leaving an unbroken droplet behind the contact line at t=1148 ms. 

A fluorescence dye of Nile red at 5 $\mu$M was added into octanol to trace the flow of octanol indicated by the bright part in the image. Figure \ref{CM} (b) shows the fluorescence microscope image of the octanol droplet with radius of 139 $\mu$m and with v$_{cl}$ of 0.9 mm/s. Similar with Figure \ref{CM}(a), the octanol droplet is completely left behind the contact line of the water drop. This stable behaviour of oil droplet, i.e., being completely transferred from water to air, is dominated by capillary force, which we will discuss later.

The octanol droplet with radius of 419 $\mu$m exhibits radically contrasting behaviour at v$_{cl}$ of 0.07 mm/s as shown in Figure \ref{CM} (c). The contact line passes the edge of microdroplet, but does not sweep cross the entire droplet, as shown in Figure \ref{CM} (c, 6 ms to 20 ms). In 60 ms, octanol in the microdroplet spreads along the contact line and evolves into a crescent shape like a bow. The crescent is darker than water, as shown in Figure \ref{CM} (c, 80 ms to 400 ms). Figure \ref{CM} (d) shows the similar spreading phenomenon captured by using fluorescence microscopy where the radius of the octanol droplet was 251 $\mu$m and the v$_{cl}$ is 0.01 mm/s. At 300 ms, a bow shape of octanol microdroplet appears in the image, which reflect the spreading of the microdroplet along the deformed contact line as a result of the Marangoni stress.

Figure \ref{CM} (e) show the typical snapshots of a microdroplet with the radius of 145 $\mu$m at v$_{cl}$ of 0.01 mm/s. In the images taken by the laser scanning confocal microscope, the red, green and black colors indicate octanol, water and air phases, respectively. We see that the contact line moves over the octanol droplet once the contact line meets the edge of the microdroplet, as shown in Figure \ref{CM} (e, 0 ms to 20 ms). After that, the octanol spreads laterally along the contact line and gradually evolves to a bow shape, as shown in Figure \ref{CM} (e, 40 ms to 160 ms). Simultaneously, a part of water was stranded on the substrate due to the pinning effect. 

Tiny twin water droplets were left on lateral sides of the octanol microdroplet, as shown in Figure \ref{CM} (e, 160 ms). Finally, octanol fully cloaks the water-air interface and shapes like a circle, as shown in the left image in Figure \ref{CM} (f). The spreading behaviour shown in Figure \ref {CM} (c-f) is dominated by Marangoni effect, which we will discuss later. The dynamics of contact line before and after the contact with the edge of microdroplet is sketched in Figure \ref{CM} (f). 

\subsection{Transition from capillarity-dominated to Marangoni effect-dominated regime}

\begin{figure}
  \centerline{\includegraphics[width=0.82\linewidth]{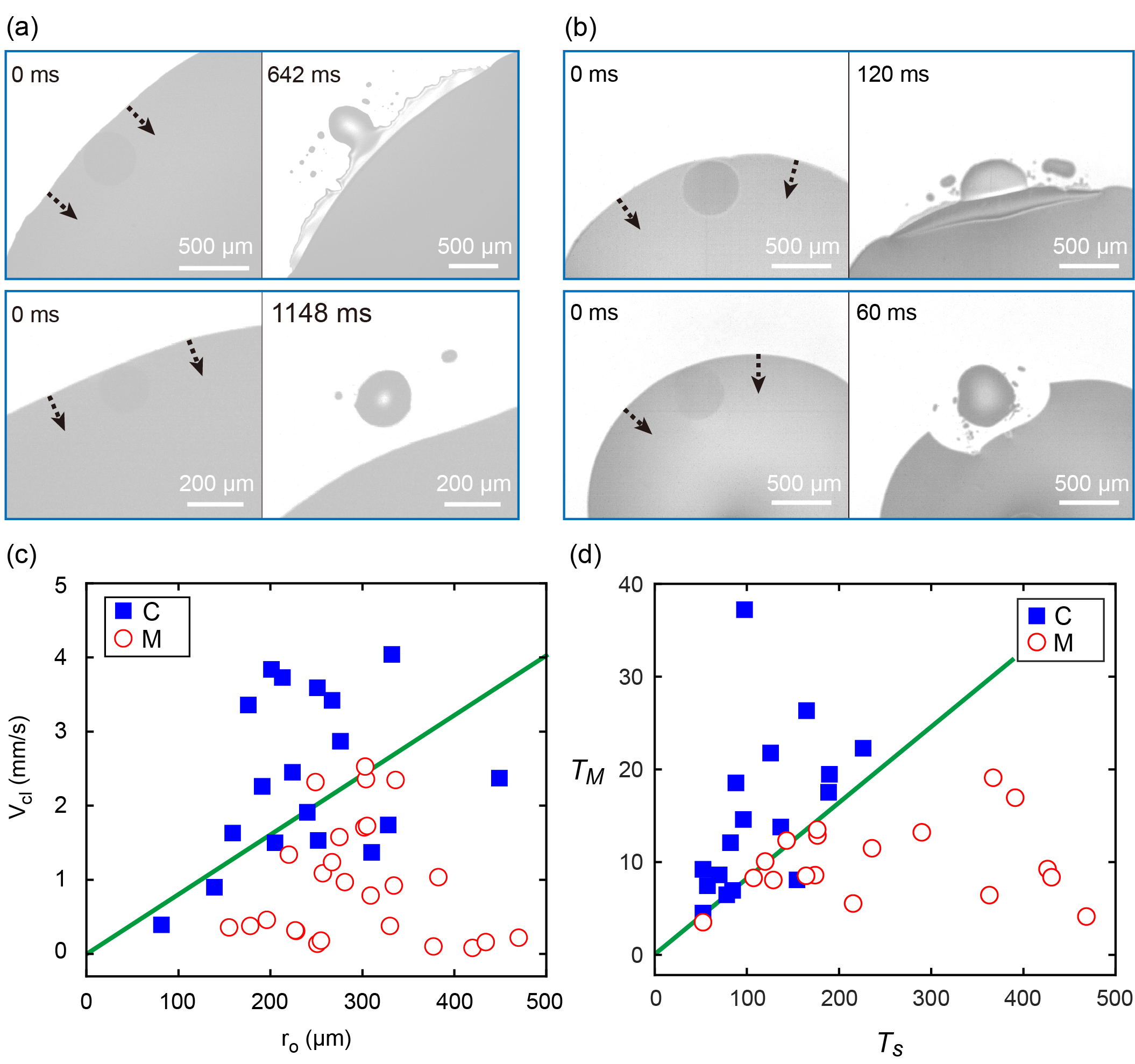}}
  \caption{Experimental results for octanol droplet encapsulated in water drop (C: capillarity-dominated; M: Marangoni effect-dominated). (\textit{a}) Up: Marangoni effect-dominated spreading behavior when $r_o$=193 $\mu$m and $v_{cl}$=0.36 mm/s; Down: capillarity-dominated stable behavior when $r_o$=83 $\mu$m and $v_{cl}$=0.36 mm/s. (\textit{b}) Up: Marangoni effect-dominated spreading behavior when $r_o$=227 $\mu$m and $v_{cl}$=0.32 mm/s; Down: capillarity-dominated stable behavior when $r_o$=213 $\mu$m and $v_{cl}$=3.73 mm/s.(\textit{c}) Phase diagram between velocity of contact line vs radius of octanol droplet. (\textit{d}) Phase diagram between time scale for contact line sliding vs time scale for oil spreading.} 
\label{transit}
\end{figure}

By varying the values of $r_{o}$ and $v_{cl}$, we are able to identify the transition between spreading (Marangoni effect-dominated) and stable (capillarity-dominated) behaviors. In Figure \ref{transit} (a), the octanol microdroplet spreads along contact line when $r_{o}$=193 $\mu$m and $v_{cl}$=0.36 mm/s, while the octanol microdroplet remains stable when $r_{o}$=83 $\mu$m and $v_{cl}$=0.36 mm/s. This result indicates the smaller $r_o$ may lead to the transition from Marangoni effect-dominated behavior to capillarity-dominated behavior. In Figure \ref{transit}(b), the octanol microdroplet spreads along the contact line when $r_{o}$=227 $\mu$m and $v_{cl}$=0.32 mm/s, but remains stable when $r_{o}$=213 $\mu$m and $v_{cl}$=3.73 mm/s. This indicates that the increase in  $v_{cl}$ may also lead to the transition from Marangoni effect-dominated behavior to capillarity-dominated behavior.

To get more insight into the dynamics of this four-phase system, we conducted systematic experiments by tuning the $v_{cl}$ from 0.01 mm/s to 5 mm/s and $r_o$ from 80 $\mu$m and 500 $\mu$m. Figure \ref{transit}(c) shows the phase diagram of $r_o$ versus $v_{cl}$. Clearly, there is a demarcation line between capillarity-dominated and Marangoni effect-dominated behavior. Smaller $r_o$ and faster $v_{cl}$ contribute to the transition from Marangoni effect-dominated behavior to capillarity-dominated behaviour. 

Next, we seek to confirm that the spreading behaviour is driven by Marangoni effect. After formation of four-phase contact line, the original water-air interface is gradually displaced by water-octanol and octanol-air interfaces. The Marangoni effect driven spreading can be identified by the spreading coefficient $S = \delta_{12} - (\delta_{13} + \delta_{23})$, where $\delta_{ij}$ is interfacial tension between phase $i$ and phase $j$ \cite{harkins1952physical}. The fluid of phase $3$ will spontaneously spread between phase $1$ and phase $2$ if $S > 0$. In our case, the phase $1$, $2$, $3$ are water, air and octanol,  respectively. The  spreading coefficient $S$ can be calculated by 
\begin{equation}
S = \delta_{wa} - \delta_{ow} - \delta_{oa} = 36.9 (mN/m),
\label{S}
\end{equation}
where $\delta_{wa}$, $\delta_{ow}$, and $\delta_{oa}$ are interfacial tensions between water-air interface (72.8 mN/m), water-octanol interface (8.4 mN/m), and octanol-air interface (27.5mN/m), respectively.

To get a better understanding of this Marangoni effect driven spreading behavior, the key is to reveal the coupling between the Marangoni effect and the sliding of water drop contact line acting on the octanol microdroplet. The time scale for the contact line sliding through the oil droplet is: 
\begin{equation}
\tau_s \sim r_{o}/v_{cl}.    
\end{equation}
The time scale for the oil droplet spreading on the water drop can be obtained by
\begin{equation}
\tau_M \sim \sqrt{\frac{\rho_w r_w^3}{\Delta \sigma}},
\end{equation}
where $\rho_w$ is the water density, and $\Delta \sigma$ is the interfacial tension difference between octanol-water interface and water-air interface. 

For the oil droplet spreading on the water drop, sliding time scale should be longer than spreading time scale, $\tau_s > \tau_M$. Figure \ref{transit}(d) displays the phase diagram replotted with $\tau_s$ vs $\tau_M$. It clearly shows that two phases are separated by a straight line, below which the octanol droplet always spreads. The slope of the straight line is measured to be about 0.09 in the range of 0 $\lesssim \tau_{M} \lesssim$ 40. Notably that the measured slope is not equal to 1, which is due to the time scales of $\tau_s$ and $\tau_M$ only standing for the order of magnitude, but not the exact values.

\begin{figure}
  \centerline{\includegraphics[width=0.85\linewidth]{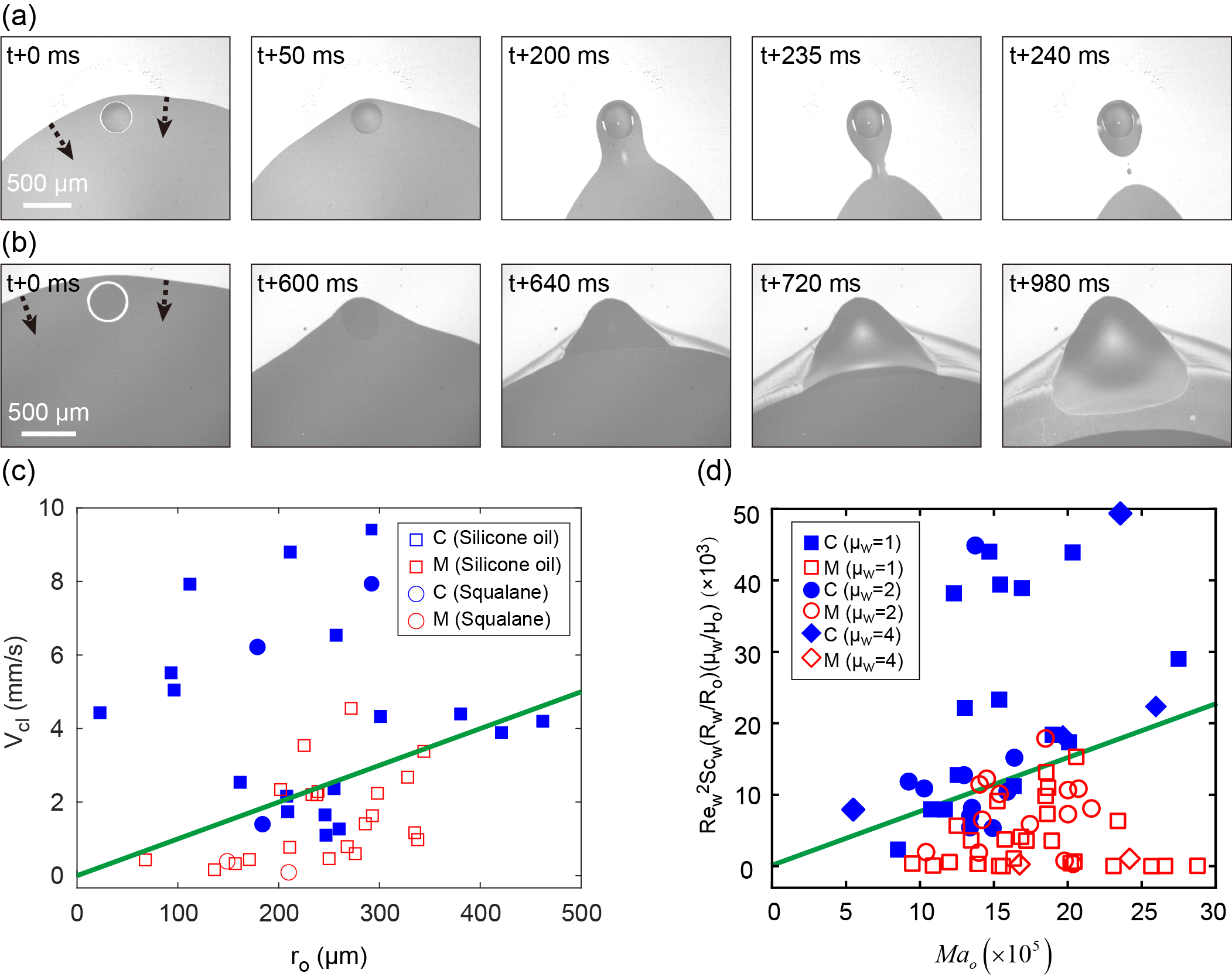}}
  \caption{Experimental results for silicone oil droplet encapsulated in water drop (C: capillary; M: Marangoni effect). (\textit{a}) Capillarity-dominated behaviour when $r_o$=161 $\mu$m and $v_{cl}$=2.54 mm/s. (\textit{b}) Marangoni effect-dominated when $r_o$= 225 $\mu$m and $v_{cl}$=3.57 mm/s. (\textit{c}) Phase diagram of capillarity-dominated and Marangoni effect-dominated behavior in the parameter space of $V_{cl}$ vs $r_o$. (\textit{d}) Phase diagram between two dimensionless numbers for the octanol microdroplet being encapsulated in water-glycerol drop with different viscosity.} 
\label{viscous}
\end{figure}

When we use silicone oil or squalane, the magic wand work well with same principle, as shown in Figure \ref{viscous}(a-c). Similar with octanol oil, silicone oil and squalane droplet also transit from  Marangoni effect-dominated behaviour to capillarity-dominated behaviour when decreasing $r_o$ and increasing $v_{cl}$.

Here we perform the dimensionless analysis for the spreading droplet. The Reynolds number $\rm{Re_w}$ is defined as the ratio of the inertial force and viscosity force on the water drop. The Schmidt number $\rm{Sc_w}$ is defined as the ratio of kinematic viscosity and mass diffusivity. 

\begin{equation}
Re_w = \rho_w v_{cl} r_w/\mu_w
\label{Re}
\end{equation}

\begin{equation}
Sc_w = \mu_w/(\rho_w D)
\label{Sc}
\end{equation}

The Marangoni number $\rm{Ma_o}$ can be used to estimate the significance of the Marangoni effect acting on the octanol microdroplet. 

\begin{equation}
Ma_o = r_{o} \Delta \sigma/(\mu_o D)
\label{Ma}
\end{equation}

Here $r_w/r_o$ is the ratio of water drop radius and octanol droplet radius; 
$\mu_w/\mu_o$, the ratio of the dynamic viscosity of water and octanol, and $D$, the mass diffusivity of octanol in water. Then $ \tau_M > \tau_s$ can be rewritten as

\begin{equation}
Ma_o > Re_w^2 Sc_w (\frac{r_w}{r_o})(\frac{\mu_w}{\mu_o}).
\label{criteria2}
\end{equation}

To demonstrate the robustness of dimensionless criteria, $\mu_w$ was tuned within the range of 1.0 mPa$\cdot$s $< \mu_w < $ 4.0 mPa$\cdot$s by adding glycerol into the water drop. By substituting the corresponding $\mu_w$, $\mu_o$, $\Delta \sigma$, $\rho_w$, $r_o$, $r_w$ and $v_{cl}$ depicted in Tables \ref{table1} and \ref{table2}, we obtain the plot of $Re_w^2 Sc_w (r_w/r_o) (\mu_w/\mu_o)$ vs $Ma_o$, as shown in Figure  \ref{viscous}(d). It clearly shows two phases are separated by a straight line and the spreading takes place when $Ma_o$ is below the reference line. 

\begin{table*}
\centering
\small
  \caption{\ Physical properties of oil in microdroplets. Note:  $\delta_{ow}$ and $\delta_{oa}$ are interfacial tensions between oil-water interface, and oil-air interface, respectively.}
  \label{table1}
  \begin{tabular*}{0.85\textwidth}{@{\extracolsep{\fill}}lllll}
    \hline
    Oil types & Viscosity  & $\delta_{ow}$ & $\delta_{oa}$  & Contact angle  \\
      &  (mPa$\cdot$s) &  (mN$\cdot m^{-1}$) &  (mN$\cdot m^{-1}$) & in water ($^\circ$) \\   
    \hline
      1-octanol     & 7.4    & 8.4      & 27.5   &  34 \\
      silicone oil  & 48.4   & 35.0     & 18.85  &  120 \\
    \hline
  \end{tabular*}
\end{table*}

\begin{table*}
\centering
\small
\caption{\ Experimental conditions for 1-octanol.}
  \label{table2}
  \begin{tabular*}{0.85\textwidth}{@{\extracolsep{\fill}}lllll}
    \hline
    water solution  viscosity          &   $r_o$    &   $r_w$  &   $r_{cl}$ \\   
    (mPa$\cdot$s)  &   ($\mu$m)   &   (mm)     &   (mm/s) \\
    \hline
       0.98                    &   50-600           &   0.8-7.0        &   0.01-7.0   \\
      2.00                    &   50-600           &   1.0-4.0        &   0.01-7.0   \\
      4.05                    &   50-500           &   1.0-4.0        &   1.0-3.0   \\

    \hline
  \end{tabular*}
\end{table*}

\subsection{Microdroplet shaping under Marangoni stress}

\begin{figure}
  \centerline{\includegraphics[width=0.85\linewidth]{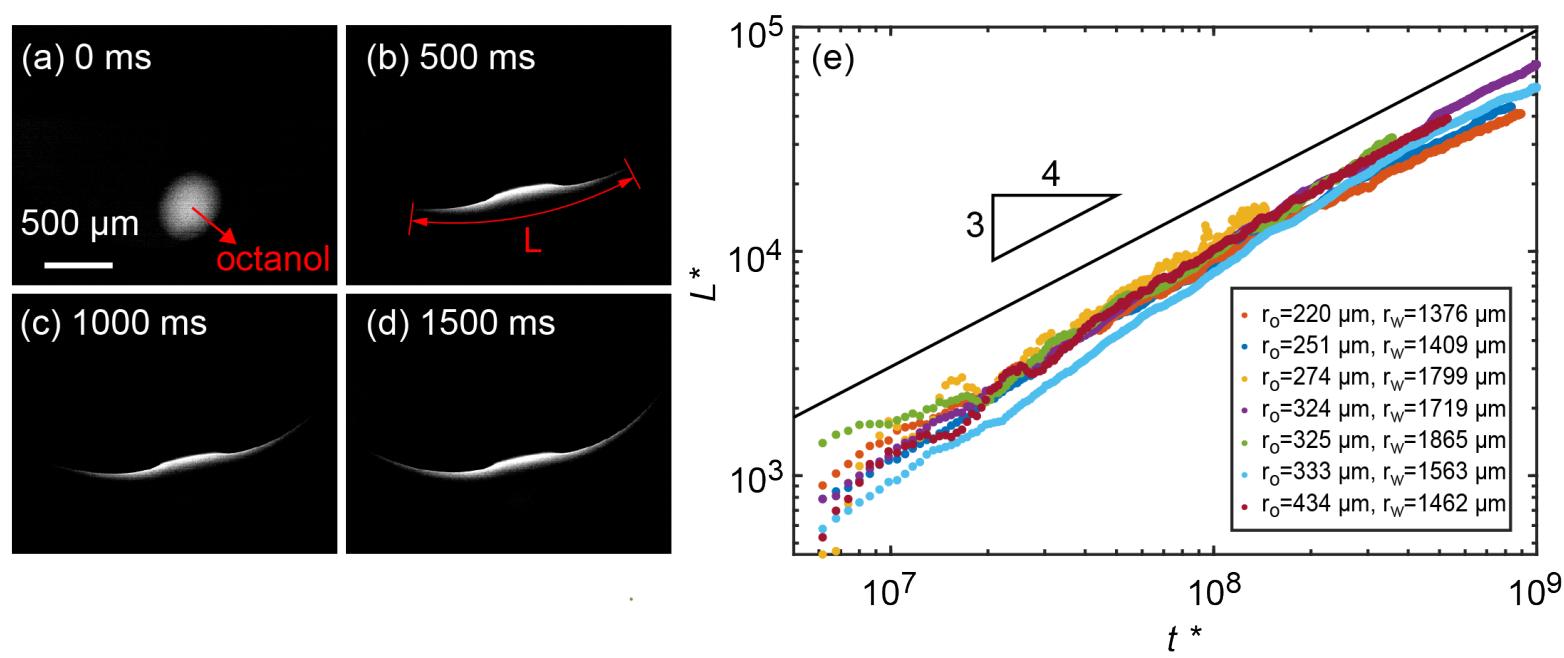}}
  \caption{Quantitative analysis of spreading length with time for octanol spreading along solid-water-air interface. Normalized spreading length $L^*$ as a function of normalized time $t^*$ in seven different tests with various octanol microdroplet size and various water drop size. $r_o$ and $r_w$ represent the radius of oil microdroplet and of water drop, respectively, and $v_{cl}$$<$ 0.1 mm/s for all the seven tests. The black solid line is the reference line of slope of 3/4. All the data come from fluorescence microscope experiments.} 
\label{arc}
\end{figure}

Next, we quantitatively investigate the evolution of spreading length of the microdroplet, $L(t)$, with time $t$. Figure \ref{arc}(a$-$d) show the typical snapshots captured by fluorescence microscope with $r_o$ = 251 $\mu$m and $r_w$ = 1409 $\mu$m. We observed that octanol phase continues extending to lateral sides with time, and forms an arc shape. Here we define the arc length $L(t)$ to represent the boundary of the drop covered by oil, as shown in Figure \ref{arc} (b). 
 
In Figure \ref{arc}(e), we show the normalized spreading length $L^\ast$ as a function the normalized time $t^\ast$. 

\begin{equation}
L^\ast = L/[\eta_1^2/(\Delta\delta\cdot\rho_1)]
\label{arcL}
\end{equation}
 
\begin{equation}
t^\ast = t/[\eta_1^3/(\Delta\delta^2\cdot\rho_1)]
\label{time}
\end{equation}

where $\eta_1$ = 1.0 mPa$\cdot$s is dynamic viscosity of water, $\Delta\delta$ = 45.3 mN/m is the difference in interfacial tensions between water-air interface and octanol-air interface, and $\rho_1$ = 998 kg/m$^3$ is the density of water. 

All the data in Figure \ref{arc}(e) were extracted from snapshots of fluorescence microscope experiments, with $r_o$ changing from 220 to 434 $\mu$m, $r_w$ changing from 1376 to 1865 $\mu$m and the velocity of water drop contact line $v_{cl}$ less than 0.1 mm/s. We find that $L(t)$ increases with time, and all seven sets of data collapse to one curve and are parallel to the reference solid line, corresponding to a scaling law of $L^\ast \sim (t^{\ast})^{3/4}$.

According to literature \cite{lohse2020physicochemical,koldeweij2019marangoni}, for Marangoni dominated spreading, the spreading length $L$ obeys a scaling law of \emph{$L \sim t^{3/4}S^{1/2}/(\rho_w \eta_w)^{1/4}$}, where $\rho_w$ is density of water and $\eta_w$ is viscosity of water. The consistency of this scaling law with our experimental results, further demonstrates that the spreading behavior of octanol along the water-air interface is governed by the Marangoni stress.

\subsection{Shaping viscous microdroplets by the receding contact line}

\begin{figure}
\centerline{\includegraphics[width=0.82\linewidth]{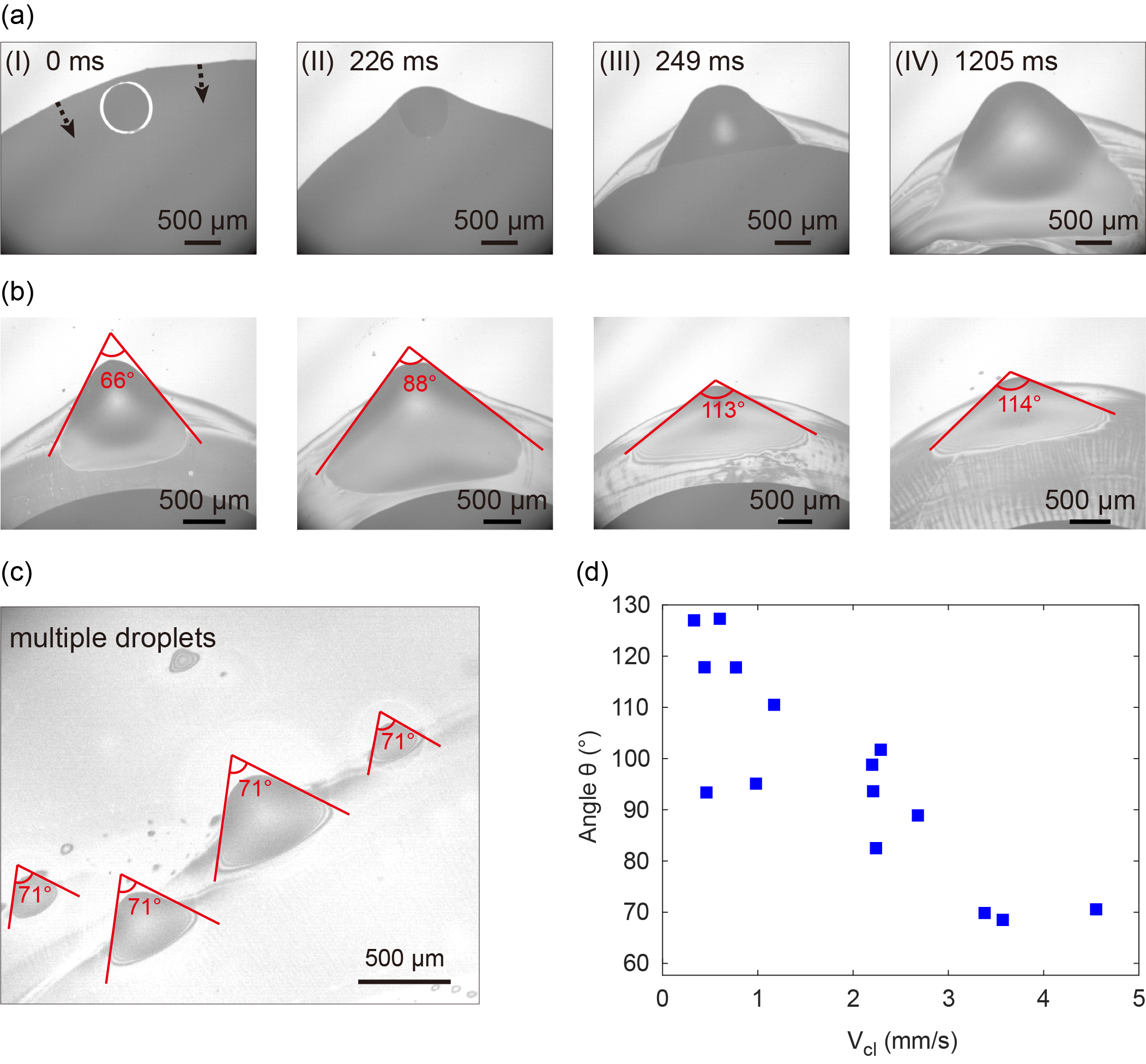}}
  \caption{Experimental results for silicone oil droplet encapsulated in water droplet. (\textit{a}) Snapshots for contact line passing through a silicone oil droplet. (\textit{b}) Different vertex angles of the silicone oil droplet at different $V_{cl}$, they are 3.57 mm/s, 2.68 mm/s, 0.77 mm/s and 0.44 mm/s, respectively. (\textit{c}) Snapshot of multiple silicone oil droplets after contact line passing through. (\textit{d}) Vertex angle of the silicone oil droplet after contact line passing through as a function of velocity of contact line. } 
\label{angle}
\end{figure}

When we changed the oil type from octanol to silicone oil, the viscosity, as well as interfacial tension, were changed. As expected, the phenomenon was different from octanol droplets. As shown in Figure \ref{angle} (a-I, a-II) the contact line of the water drop was pinned by the silicone oil droplet, rather than sliding immediately on the oil drop after the water contact line contact with oil drop. At one moment, as shown in Figure \ref{angle} (a-III), the contact line of water drop depins and slides quickly on the oil droplet. At the same time, the oil spreads along water-air interface under the Marangoni stress. In this process, the shape of the silicone oil droplet evolves into a shape that is broad, triangular, and have smoothly rounded edges, resembling the shape of the pectoral fins of a stingray. Next, we conducted the experiments with multiple silicone oil droplets encapsulated in a water drop, all the silicone oil droplets are shaped into similar shapes, as shown in Figure \ref{angle} (c). Changing the velocity of water-air contact line ($v_{cl}$), we found a negative relationship between vertex angle of the triangle and $v_{cl}$, as shown in Figure \ref{angle} (d). This indicates that we can shape viscous oil droplet by changing the moving velocity of water contact line. On the other hand, the behaviour of viscous liquid drop under the Marangoni stress implies that those viscous microdroplets may not be removed from the surface during cleaning. 

\section{Conclusion}
A sessile oil microdroplet may spread or remain stable when displacing the surrounding phase from an immersing aqueous drop to air. The morphology of the microdroplet transferred into the air is determined by the interfacial dynamics at the four-phase contact line of air-solid-water-oil. The oil microdroplet spreads along the contact line under the Marangoni stress at slower $v_{cl}$ (moving velocity of contact line) and larger $r_o$ (radius of oil microdroplet) with the spreading distance following a scaling law of 3/4 with time. The oil microdroplet remains stable, being transferred from water phase to air at a faster $v_{cl}$ and smaller $r_o$ due to the capillary force. The transition from the Marangoni effect-dominated regime to the capillarity-dominated regime is attributed to the competition between $\tau_s$ (the time scale for water-air contact line moving through the oil microdroplet) and $\tau_M$ (the time scale for the oil microdroplet spreading on water drop). Without spreading out, a viscous oil microdroplet adapts to a characteristic shape in response to the Marangoni stress at the contact line. The insights from this work may help manipulate the oil droplets between immiscible phases in the applications for open microfluidics, droplet-based extraction, decontamination and surface cleaning.

% Experimental section
\section{Experimental Section}
\threesubsection{Chemicals and Materials}1-octanol (90\% Fisher) and silicone oil (Aldrich) were used as received. Water (18.2 M$\Omega\cdot$cm) were used as received. Glycerol (Fisher) was used to adjust the viscosity of water without the significant change in surface tension \cite{takamura2012physical}. Water soluble fluorescein isothiocyanate-dextran (Sigma-Aldrich) and oil soluble nile-red (Sigma-Aldrich) were used in confocal microscope experiments to label water and octanol, respectively. 3-aminopropyl triethoxysilane (APTES) (98\%, Alfa Aesar) was used for the surface modification of the glass slides.

\threesubsection{Substrate preparation}High-precision cover glasses with three dimensional size of 60 $\times$ 24 $\times$ 0.17 $mm^3$ was hydrophobilized with APTES by the following procedures. The slides were first ultrasonically cleaned by using water and ethanol in sequence for 10 min each, and then were dried in a stream of air. The pre-cleaned slides were further washed in piranha solution (30 vol\% hydrogen peroxide and 70 vol\% sulfuric acid) at 80 $^{\circ}$C for 20 min, and then were thoroughly rinsed with water. These cleaned slides were completely dried under the air stream and then in an oven with 120 $^{\circ}$C for 1 hour. Then the slides were immersed for 12 h into a 2.0 vol\% APTES in toluene (99.5\% Fisher). Next, the slides were ultrasonically rinsed in sequence with toluene, ethanol and water for respective 10 min. Then, these freshly prepared slides were baked at 120 $^{\circ}$C for 2 h and the hydrophobization of slides was finished. Before use, the hydrophobized slides were washed in ethanol and water by ultrasonication for 10 min, respectively, and then dried under an air stream.

\begin{figure}
  \centerline{\includegraphics[width=0.82\linewidth]{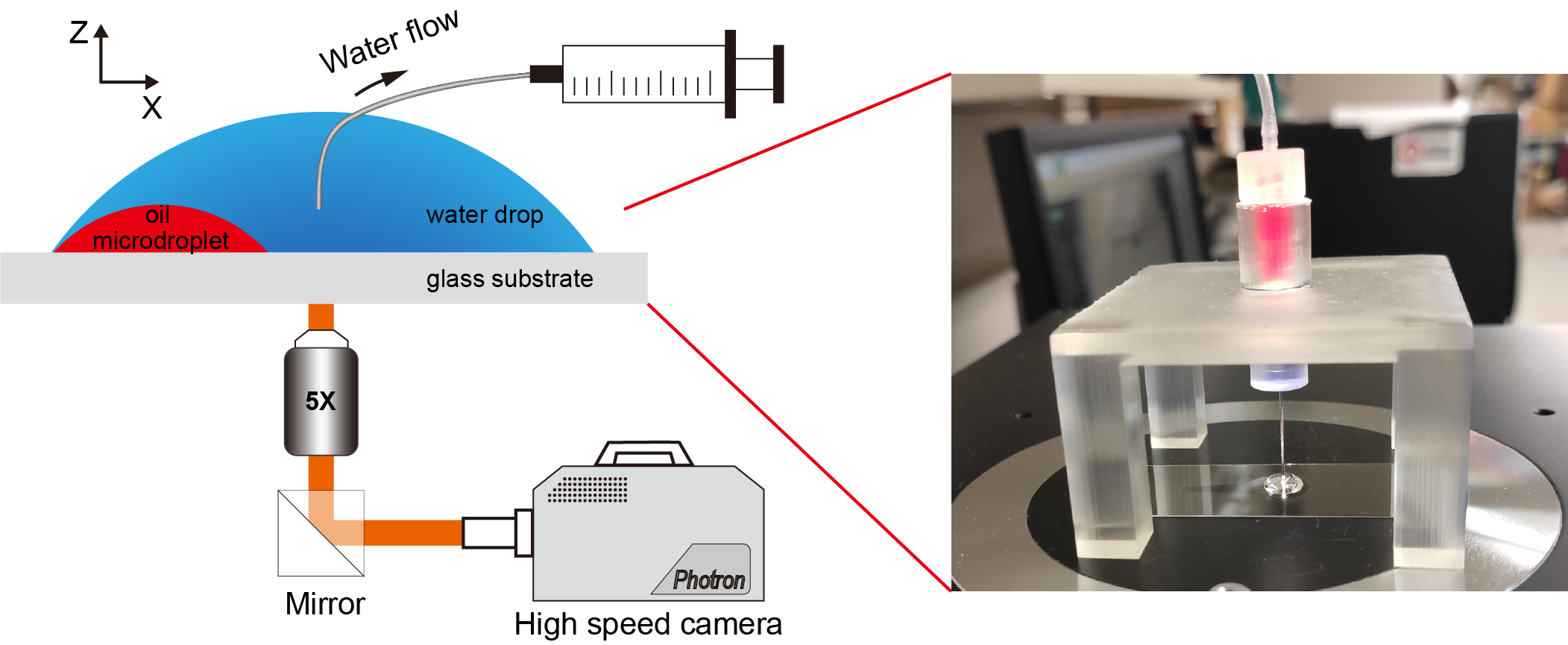}}
  \caption{Experimental setup and typical results for octanol droplet in water. (a) The schematic diagram of the experimental setup in front view; (b) Photo of the 3D printed frame for controlling receding velocity of water drop.}
\label{exp}
\end{figure}

\threesubsection{Experimental Setup}
Figure \ref{exp}(a) shows the schematic of the experimental setup. All experiments were carried out at a constant temperature of 21 $^{\circ}$C. A small 1-octanol microdroplet was firstly placed on the APTES substrate by using a microsyringe. Then a water drop was directly deposited on the 1-octanol droplet. The water drop was much larger than the 1-octanol microdroplet, completely covering the 1-octanol microdroplet. An injection pump was employed to remove the water droplet with a steady flow rate, making the contact line of the water drop to move towards the 1-octanol microdroplet. A high speed camera (Photron HX100, Japan) equipped with a 5× objective (Olympus, Japan) was used to capture the dynamics of these two droplets from the bottom view with frame rate from 1 kHz to 3.2 kHz. The relevant physical and chemical properties of substrate and chemicals and the experimental conditions are listed in Table \ref{table1} and Table \ref{table2}, respectively.

The velocity of contact line is defined as the instantaneous velocity when three-phase contact line is getting closer to oil droplet. When we need different velocities, we can control the flow rate by the injection pump, and the expectant velocity can be calculated through the flow rate, the water drop size and the deviation of oil droplet from the center of the water drop.

\threesubsection{Confocal Microscopic imaging}5 $\mu$M of FITC-Dextran and 5 $\mu$M of Nile-red were respectively employed to label water and 1-octanol. The injection pump was not used here. The contact line receding of the water droplet was realized by the evaporation of water. Laser scanning confocal microscopes (Leica Stellaris 5) equipped with 10× objective were employed to capture the dynamic process with frame rate of 50 Hz. 488/534 nm laser beams were respectively used to excite the FITC-Dextran and nile-red in the droplets.

\section{Acknowledgements}
We are grateful for the inspiring and critical comments from Professor Detlef Lohse at University of Twente. H.C.Y. appreciates China Scholarship Council for the financial support. X.H.Z acknowledges the funding support from the Natural Science and Engineering Research Council of Canada (NSERC)- Discovery Project, NSERC Alliance – Alberta Innovates Advance grants, the Canada Research Chair Program and Canada Foundation for Innovation, John R. Evans Leaders Fund. B.X. is grateful for the support from the Engineering and Physical Sciences Research Council (EPSRC, UK) grant-EP/N007921 and EP/X032582.

\medskip
\textbf{Supporting Information} \par 
The videos showing the process in Figure \ref{CM} (a-e) are attached as supporting information. The file 'video\_a.avi' displays the process in Figure \ref{CM} (a). The file 'video\_b.avi' displays the process in Figure \ref{CM} (b). The file 'video\_c.avi' displays the process in Figure \ref{CM} (c). The file 'video\_d.avi' displays the process in Figure \ref{CM} (d). The file 'video\_e.avi' displays the process in Figure \ref{CM} (e). 

Supporting Information is available from the Wiley Online Library or from the author.

\medskip
\textbf{Declaration of interests} The authors declare that they have no known competing financial interests or personal relationships that could have appeared to influence the work reported in this paper.

\medskip
\textbf{Author contributions}
HCY and BLZ contributed equally to this work. BBX and XHZ: Conceptualization. HCY and BLZ: Methodology, data collection and analysis. HCY, QYL, BBX and XHZ: Writing original draft, editing and review. YWX,XHG,YJC: Project administration and funding acquisition.

% References
\medskip
\bibliographystyle{MSP}
\bibliography{Ref}

% Table of contents entry should be 50 - 60 words long
% Image should be 55 mm broad and 50 mm high or 110 mm broad and 20 mm high

\end{document}